\documentclass[a4paper]{article}

\usepackage{epsfig}% Include figure files
\usepackage{graphicx}
\usepackage{epsfig}
\usepackage{dcolumn}% Align table columns on decimal point
\usepackage{bm}% bold math
\usepackage{amsmath}
\usepackage{amssymb}
\usepackage{amsfonts}
\def\expect{\mathbb E}
\def\vdeg{\Lambda}

\def\evdeg{\lambda}
\def\efdeg{\rho}
\def\uy{\underline{y}}
\def\ux{\underline{x}}
\def\ind{{\mathbb I}}
\begin{document}

\title{Statistical Physics of Group Testing}
\author{Marc M\'ezard \thanks{LPTMS, Universi\'e Paris-Sud, UMR8626, B\^at. 100, 91405
Orsay cedex, France}, Marco Tarzia \thanks{LPTMS, Universi\'e Paris-Sud, UMR8626, B\^at. 100, 91405
Orsay cedex, France} and Cristina Toninelli \thanks{LPMA, CNRS-UMR 7599, Univ. Paris VI-VII, 4 Pl. Jussieu, 
Paris, 75005 France}}
\maketitle
%\date{\today}

\begin{abstract}
This paper provides a short introduction to the group testing problem, and reviews various aspects of its statistical
physics formulation. 
%C To 
Two main issues are discussed: the optimal design of pools used in a two-stage testing experiment, like the one often used in medical or biological  applications, and the inference problem of detecting defective items based on pool diagnosis. The paper is largely based on:
M. M\'ezard and C. Toninelli, arXiv:0706.3104i, and 
M. M\'ezard and M. Tarzia {\it Phys. Rev. E} {\bf 76}, 041124 (2007).
\end{abstract}

%\maketitle

\section{Introduction}
Group testing dates back to 1943, when Dorfman suggested to use it for testing
whether US draftees had syphilis~\cite{Dorfman}. Instead of testing each
individual blood sample, the idea is to mix the blood and test pools. In a
group of $N$ soldiers, one can first test $N/k$ pools of $k$ individuals. Then
one focuses on the infected pools and performs a second stage of tests on all
the individuals belonging to these infected pools. Assuming that each soldier
is infected with probability $p$
%C has the probability $p $ to be infected 
(and that the infections are
uncorrelated), the total expected number of tests is
\begin{equation} 
\expect T = { \frac{N}{k}} +{
\left[1-(1-p)^k\right]\frac{N}{k} k}\ .
\end{equation}
Minimizing this expression over $k$, one finds that the optimal size of the
pools is $k\sim1/\sqrt{p}$, giving an expected number of tests $\expect
T\simeq 2 \sqrt{p} N $. If the prevalence of infection is small, $p \ll 1$,
Dorfman's proposal reduces the total number of blood tests, compared to the
individual tests, by a factor $2\sqrt{p}$.

 The general problem of group testing~\cite{book} is that of identifying
 defectives in a set of items by a series of tests on pools of items,
 where each test only detects whether there exists or not 
%C a defective
at least a defective item in
 the pool. It has numerous applications. In particular it is being used in
 building physical maps of the genome, by detecting whether a special target
 subsequence of bases is present in a DNA strand~\cite{clone1,clone2}.
 But it has also
 been suggested to use it in HIV detection~\cite{Zenios}, in detecting failures in
 distributed computation~\cite{rish},  or in data
gathering in sensor networks~\cite{sensor}.

We shall subdivide the group testing problem into two main topics: the pool
design (building optimal pools, exploiting the possibility to have overlapping
tests), and the inference problem (how to detect defective items, given the
results of the pool tests). Two other important classification patterns are
the number of stages of detection, and whether one needs to be sure of the
result (assuming that the tests are perfect). For instance Dorfman's design is
a two-stage algorithm with sure result. In a first stage one tests the $N/k$
pools, and based on the results of this first stage one designs the second
stage of tests, namely the list of all individuals whose blood sample belonged
to a defective pool in the first stage. Getting sure results (in contrast to
results that hold with high probability) is often needed in medical
applications. In order to obtain them, one ends a group testing procedure by a
final stage (the second stage in Dorfman's procedure) which tests individually
all the items for which the procedure did not give a sure answer.

In all our study we suppose that the status ('defective' or 'OK') of all the $N$
items under study are iid random variables: each item can be defective with a
probability $p$, or OK  with probability $1-p$. Furthermore we assume that the
value of $p$ is known. This framework is called 'probabilistic group testing'
in the literature. Another framework has been studied a lot, that of
combinatorial group testing where the number of defective items is supposed to
be known. Reviews on these two frameworks can be found in~\cite{book,review1}.

Clearly one can detect defectives more efficiently (with less tests) when more
stages can be done. How efficient can one be? Information theory provides an easy lower bound, which applies to the situation where there is no limit on the allowed number of stages.  Assume you have a total of $N$ items, and you
perform  $T$ tests altogether. There are $2^T$
possible outcomes.
If there were  $d$ defectives, a necessary condition to
detect them is that $2^T \ge \binom{N}{d}$. In the large $N$ limit, 
taking  $d=N p$, one finds that the number of tests $T$ must be larger than:
\begin{equation}
T \ge N H_2(p) = N\left(-p \log_2 p -(1-p) \log_2 (1-p) \right)\ .
\end{equation}
It is not difficult to design a sequence of pools, with an unbounded number of
 stages, which basically 
reaches this limit.

Clearly, when $p$ is small, the minimal number of tests
in unbounded number of stage, $N (-p \log_2 p)$, is much smaller
than Dorfman's two-stage result $2 N\sqrt{p}$. A natural question
is that of the minimal number of tests, and the corresponding
 best pool design, if the number of stages is limited to a value $S$.
 In the next section we give the answer when $S=2$, for the case of sure results in two-stage procedures. Amazingly there exist pool designs which 
require only $ N H_2(p) /\log 2$ tests, a factor $1.4$ larger than the
optimal unbounded-stage result.
\section{ Optimal two-stage design in the small $p$ limit}
In their  nice analysis of two-stage group testing, Berger and Levenshtein~\cite{Berger,Berger2} 
suggest to study the case where the number of items $N$ goes to infinity,
and the probability of being defective $p$ goes to zero like
$p=N^{-\beta} $. In this limit they obtain the following bounds for the 
%C perhaps better to add ``minimal'' or ``optimal'' (if not clear we can write
%''.. minimal (over all the two stage algorithms) expected number of tests required to find all the defectives, when...''
minimal 
expected number of tests required to find all the defectives in two stages, when $\beta <1$:
$ 
1/\log 2 \leq \lim_{N\to\infty}
\overline{T}(N,p) / (N p |\log p|) \leq \frac{4}{\beta}$.

The recent work of~\cite{MezTon} has derived the exact asymptotic
%C again minimal
minimal expected number of tests, and the pool design which reaches them,
when $\beta< 1/2$ as well as in the limit $p\to 0$ after $N\to \infty$. 
In order to state the results in a compact form, it is convenient to introduce 
the notation
by $\lim_{N \to \infty \vert \beta}$ for the limit where
$N$ goes to $\infty$, $p$ goes to zero, with $p=N^{-\beta}$ and $\beta>0$. 
The limit $\lim_{p\to 0} \lim_{N\to\infty}$ will be referred to  as the $\beta=0$
case.

The two main results of~\cite{MezTon} are  the following.

\noindent
 1) If $0\le \beta< 1/2$:
\begin{equation}
\lim_{N\to\infty\vert \beta}\frac{\overline{T}(N,p)}{N p |\log p|}
=\frac{1}{{(\log 2)}^2}\ .
\end{equation}
\noindent
%C {\expect T}
2) In these limits, ``regular-regular'' pools of girth $\ge 6$, with
 tests of degree $K=\log 2/p$ and variables of degree
  $L=|\log p|/\log 2$ become optimal with probability tending to one when $N \to \infty \vert \beta$. 

  This pooling design is best understood in terms of its factor graph
  representation. One builds a graph where there are two types of vertices:
  each variable is a vertex 
%C 
(represented by a circle in figure \ref{fig:undet0}), 
and each test is also a vertex
%C
(represented by a square). An edge is
  present in the graph, between variable $i$ and test $a$, whenever the
  variable $i$ appears in test $a$. The graph is thus bipartite, with edges
  only between variables and tests. The regular-regular pools correspond to
  random factor graphs, uniformly drawn from the set of graphs where every
  variable has degree $L$ and every test has degree $K$, and such that the
  girth (the size of the shortest loop) is at least $6$. The existence of such
  graph has been demonstrated by Lu and Moura~\cite{LuMoura}. Their
  construction is a bit complicated, but a simpler class of graphs has also
  been shown in~\cite{MezTon} to reach optimal performance in the asymptotic limit $p\to 0
  \vert \beta$, with $0 \le \beta < 1/2$. These are the so-called
  ``Regular-Poisson'' graphs where each variable chooses randomly the $L=|\log
  p|/\log 2$ tests to which it belongs, uniformly among all the $\binom{M}{L}$
  possible sets of $L$ tests. In such a case the degrees of the tests become
  asymptotically Poisson distributed, with mean $ N L/M$.

The idea of the proof goes in two steps: a general lower bound of combinatorial nature, 
and a detailed analysis of the previous two random pools designs, showing that their number of tests
asymptotically matches the lower bound.

\subsection{Lower bound}
\begin{figure}
\begin{center} \includegraphics[scale=0.55]{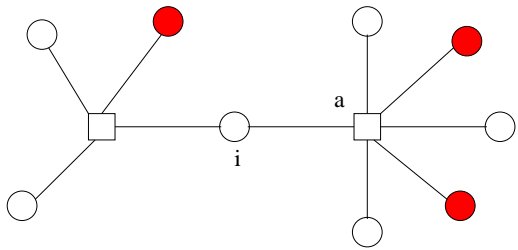}
\end{center}
\caption{ The site $i$ is an undetected $0$ whenever all the tests containing $i$ contains at least
one item $j$ which is defective (coloured here). }
\label{fig:undet0}
\end{figure}

The lower bound $\expect \; T / (N p |\log p| )\ge 1 / (\log 2)^2$
%C {\overline{T}(N,p)}
is obtained as follows. Any pool design is characterized by a graph, 
and therefore by a connectivity matrix $c$ with element  $c_{ia}=1$ if item $i$ belongs to pool $a$,
$c_{ia}=0$ otherwise. Given a graph and an item  $i$ with $x_i=0$, let us find out the condition for
this to be an undetected $0$. This situation occurs whenever any test containing $i$ contains at least
one item $j$ which is defective (see figure \ref{fig:undet0}). If the girth of the graph is larger than $6$, the values of 
the variables $x_j$ in these neighbouring checks are uncorrelated, and the expected number of undetected $0$ in the first stage, for a given graph, is
\begin{equation}
U_0= \sum_{i=1}^N\; (1-p)\; \prod_{a=1}^M \left(1-(1-p)^{d_a-1}
\right)^{c_{i,a}},
\label{eq:u0def}
\end{equation}
In general graphs (without any girth condition), one can use a Fortuin-Kasteleyn-Ginibre inequality~\cite{FKG} to show that 
$U_0$ is always a lower bound to the  expected number of undetected $0$ in the first stage.
Then  one minimizes the $U_0$ of eq.(\ref{eq:u0def}) over all graphs. This is done 
using the function $f(\vec m)$ which is the fraction of sites
such that, among its neighbouring checks, $m_1$ have degree $1$, $m_2$ have degree $2$, etc...
Both the total number of checks, and $U_0$, can be written as linear expressions in $f(\vec m)$.
Minimization over $f$ is thus easily done.

\subsection{Upper bound}
It turns out that in the small $p$ limit, the undetected $0$s are the dominant sources of errors.
One way to see this is through the study of random graph ensembles.

Imagine that the factor graph is generated from an ensemble of graphs where the 
degrees of the variables and checks are random variables drawn randomly from some fixed 
distribution. We shall adopt the usual notations from the coding community
for describing these degree sequence:

\begin{itemize}
\item
 An item has degree $\ell$ with probability $\Lambda_\ell$. The sequence of $\Lambda_\ell$ 
is encoded in the polynomial $\Lambda[x]=\sum_\ell \Lambda_\ell x^\ell$.

\item
A test has degree $k$ with probability $P_k$. 
This is encoded in $P[x]=\sum_k P_k x^k$.

\end{itemize}

It is also useful to introduce the 'edge perspective degree profiles'. Let $\lambda_\ell$
be the probability that, when one picks an edge at random in the factor graph, the variable to which it is attached has degree $\ell$, and $\rho_k$ be the probability that
the test to which it is attached has degree $k$. Then
$\lambda_\ell = \ell \Lambda_\ell/(\sum_n n \Lambda_n)$ and $\rho_k= k P_k /(\sum_n n P_n)$.
These distributions are encoded in the functions 
$\lambda[x]=\sum_\ell \lambda_\ell x^{\ell-1}$ and $\rho[x]=\sum_k \rho_k x^{k-1}$.

Assuming that we have generated a random graph with girth $\ge 6$, the various quantities that
appear in the computation of the expected number of tests can be expressed in terms of
the generating functions. 

\begin{itemize}
\item The total 
number of pools in the first stage is: $ G= N \langle \ell\rangle / 
\langle k \rangle
= N \Lambda'[1] / P'[1]$.

\item 
The number of sure OK items (girth $\ge 6$) detected in the first stage is: 
$N_0 = N (1-p)\left(1-  \vdeg\big[1-\efdeg[1-p]\big]\right)$.

\item 
Number of sure defective items (girth $\ge 6$) detected in the first stage is:
$N_1 = N p\left(1- \vdeg\Big[1-\efdeg\big[(1-p)(1-\evdeg[1-\efdeg[1-p]])\big]\Big]\right)$.

\item All the items which are not detected after the first stage must be tested individually
in the second stage. Therefore the total expected number of tests is: 
\begin{equation}
%\overline T
{\expect T}  
= G + N-(N_0+N_1).
\label{T_randgraph}
\end{equation}
This expression is to be minimized over the degree distributions $\lambda[x]$ and $\rho[x]$.
\end{itemize}

\begin{figure}
\begin{center}
\includegraphics[angle=-0,width=0.6\textwidth]{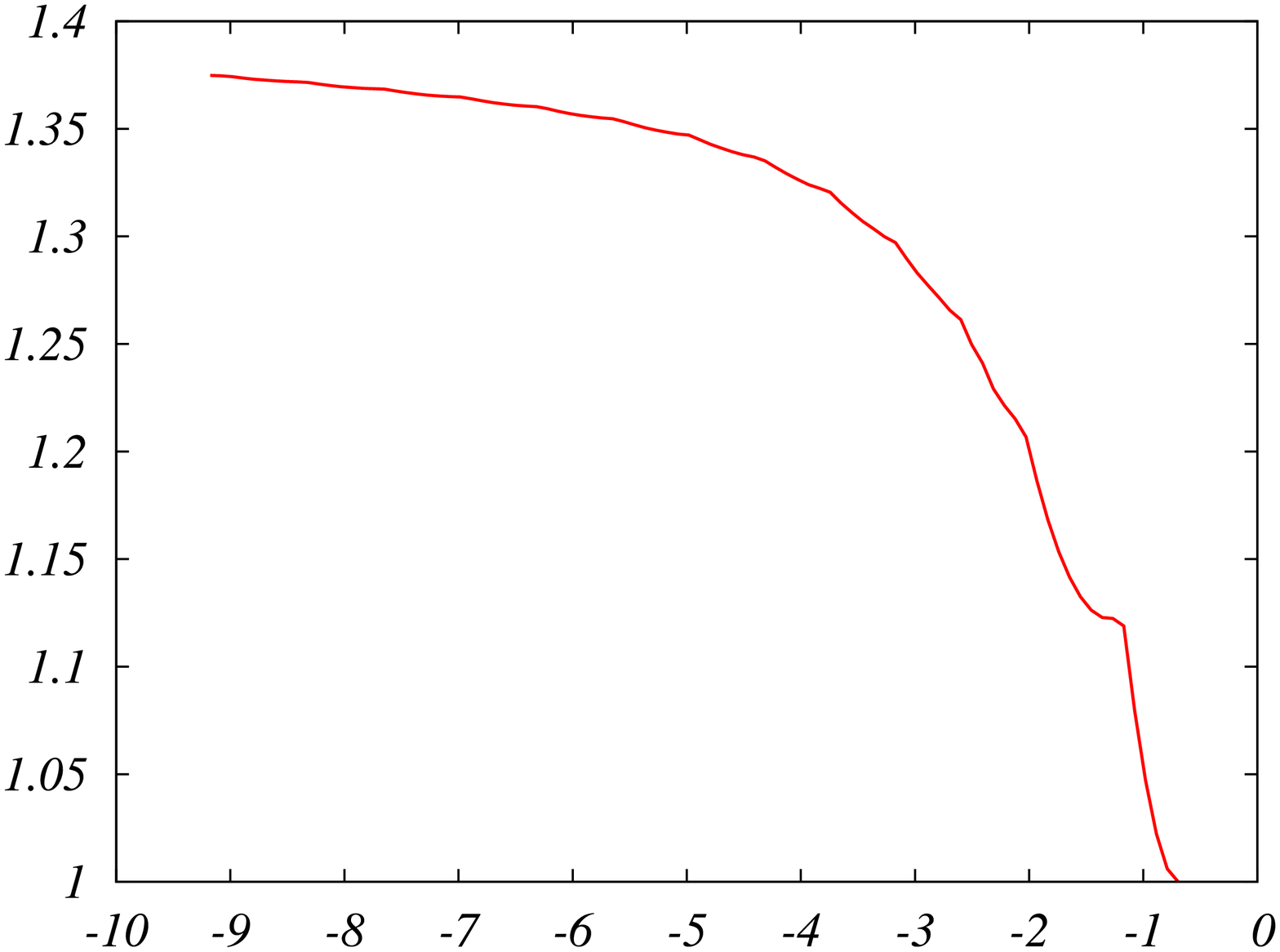}
\end{center}
\vspace{-5.0cm}
\hspace{1.3cm} {\Large $\frac{\overline{T}}{N H_2(p)}$}

\vspace{3.9cm}
\hspace{7.8cm} {\large $\log p$}
\caption{
Expected number of tests in a sure group testing with  a two stage procedure: optimal performance obtained in the optimal regular-regular random pool design. The graph gives the expected number of tests, divided by the information theoretic lower bound for arbitrary number of stage ($N H_2(p)$), plotted versus $\log p$. The non-analyticity points correspond to the values of $p$ where the optimal value of the degree pair $L,K$ changes. In the small $p$ limit
the curve goes asymptotically to $1/\log 2$.
}
\label{fig:p_finite}
\end{figure}

Some simple probability distributions can be studied efficiently. For instance, 
the regular regular one, parameterized by $\Lambda[x]=x^L$ and $P[x]=x^K$ 
%C lead
leads to a simple
result for 
${\expect T}$
%C $\overline T$ 
in (\ref{T_randgraph}). This can be optimized with respect
to $K$ and $L$ for any $p$. figure \ref{fig:p_finite} shows the result. 

When $p \to 0$, it is easy to prove that the optimal values of $K$ and $L$ are $K^*(p)= \frac{\log 2}{p}$ and $L^*(p) = -\frac{\log p}{\log 2}$, and that they saturate the previous lower bound 
%C 
thus providing the
exact asymptotic value of the minimal (over all two-stage procedures)
number of expected tests:
%C
$\overline T  =  N\left(-p \log_2 p \right)/\log 2$. This is also the case of
regular Poisson graphs.

For finite $p$, the best degree sequences $\Lambda,P$ are not known. However
we have proved that at most $3$ coefficients
$\Lambda_\ell$, and at most $5$ coefficients $P_r$ are non-zero in these optimal
sequences. Plugging this information in some numerical minimization procedure
of (\ref{T_randgraph}), we have observed numerically that for most values of
$p$ the optimal degree sequence seems to be the regular-regular one. There are
also some values where the optimal graph is slightly more complicated. For
instance for $p=.03$, the best sequences we found are $\Lambda[x]=x^4$ and
$P[x]=.45164\; x^{21}+.54836\; x^{22}$, giving 
%C $\overline T=.25450$
$\expect T=.25450$
, slightly better than the one obtained
with $\Lambda[x]=x^4$ and $P[x]=x^{22}$, giving 
%$\overline T=.25454$
$\expect T=.25454$
. But for all
values of $p$ we have explored, we have always found that either the
regular-regular graph is optimal, or the optimal graph has superposition of
two neighbouring degrees of the variables, as in this $p=.03$ case.
In any case regular-regular is always very close to the optimal structure.

\section{One stage group testing: inference}
Another interesting aspect of group testing, which we now discuss, is the 
identification of defective items, given a pool design and the results of the tests.
This amounts to minimize the number of errors  in a one stage experiment. Given a set of pools
and the corresponding tests' results, the identification of the most probable status of each variable is a typical inference problem that can be formalized as follows. Imagine that the
items status are given by  $\uy=y_1,\dots,y_N$, where $y_i=1$ if item $i$
is faulty, $y_i=0$ if it is OK. Then the test $a$ returns a signal
 $t_a=T_a(\uy)= 1$ if $\sum_{i\in V(a)} y_i \ge 1$, otherwise $t_a=0$. Given these
test results, one can compute the probability that the items status
are given by $\ux=x_1,\dots,x_N$. This is given by:
\begin{equation}
P(\ux)\ = \ \frac{1}{Z}\  \prod_{i=1}^N\left[(1-p)\delta_{x_i,0}+p\delta_{x_i,1}\right]\ 
\prod_a \ind\left[T_a(\ux)=t_a\right].
\end{equation}
Our task is to find the configuration $\ux^*$ which maximizes this probability.
The detection error will be measured by how much $\ux^*$ differs
from $\uy$.

In order to find $\ux^*$, we first use the fact that, whenever a test $a$
returns a value $T_a(\uy)=0 $, we are sure that all the variables
$i$ belonging to pool $a$ are sure 0: $y_i=0$. Therefore we know that $x_i^*=0$.
This simple remark leads us to a 'graph stripping' procedure: We can take away from the graph all tests $a$ such that $t_a=0$, and all the variables in these pools: they are sure $0$s. The remaining 'reduced graph' has only tests $a$
with $t_a=1$. With some slight abuse of notation, let us call $\ux$ the set of variables which remain in this
 reduced graph (their number is $N'\le N$). The probability distribution on these remaining variables can be written as
\begin{equation}
Q(\ux)\ =\  \frac{1}{Z'} \prod_{i=1}^{N'}\left[(1-p)\delta_{x_i,0}+p\delta_{x_i,1}\right]\ 
\prod_a \ind\left[T_a(\ux)=1\right].
\end{equation}
The reduced problem can thus be formulated as follows: Find the values of $x_1,\dots x_{N'}$ such that:
\begin{itemize}
\item For each test $a$ in the reduced graph, there is at least one of the variables 
in its pool that is defective. 
\item 
The total number of defective variables should be minimized.
\end{itemize}
This problem is a version of the celebrated  vertex cover problem~\cite{vc,weigt,weigt_zhou} to the case 
of a hyper-graph. It is known as the hitting set problem. In the next section we discuss
the statistical physics of this problem.

\section{Hitting set}
The hitting set problem is an interesting problem in itself. In order to get some experience about it, we have studied in~\cite{MezTar} the hitting set problem in case of
random regular hyper-graphs where tests have degree $K$ and variables have degree $L$.
We define the weight of a configuration as $A(\{x_i\})=\sum_{i=1}^N x_i$. 
The Boltzmann-Gibbs measure of the problem  is defined as
\begin{equation}
P(\ux)=(1/Z) e^{- \mu A (\{x_i\})}\prod_a \ind\left[T_a(\ux)=1\right].
\end{equation}
We first write the Belief Propagation (BP) equations for this problem. 
Given a graph and a variable $i$, we consider a sub-graph rooted in $i$
obtained by removing the edge between $i$ and one of its
neighbouring tests, $a$.  Define $Z_0^{(i \to a)}$ 
and $Z_1^{(i \to a)}$ as the
partition functions of this sub-graph restricted to configurations where the
variable $x_i$ is respectively OK ($x_i=0$) or defective ($x_i=1$).
If the underlying graph is a tree, these two numbers can be computed recursively as follows:

\begin{eqnarray} \label{eq:rs1}
Z_1^{(i \to a)} & = & e^{-\mu} \prod_{b \in \partial i \setminus a} 
Y^{(b \to i)}\\
Z_0^{(i \to a)} & = & \prod_{b \in \partial i \setminus a} 
Y_1^{(b \to i)} 
\\
Y_1^{(a \to i)} & = & \prod_{j \in \partial a \setminus i} 
\left( Z_0^{(j \to a)} + 
Z_1^{(j \to a)} \right) - \prod_{j \in \partial a \setminus i} 
Z_0^{(j \to a)}\\
Y^{(a \to i)} & = & \prod_{j \in \partial a \setminus i} 
\left( Z_0^{(j \to a)} + 
Z_1^{(j \to a)} \right).
\end{eqnarray}

\begin{figure}
\begin{center}
\includegraphics[scale=0.35,angle=270]{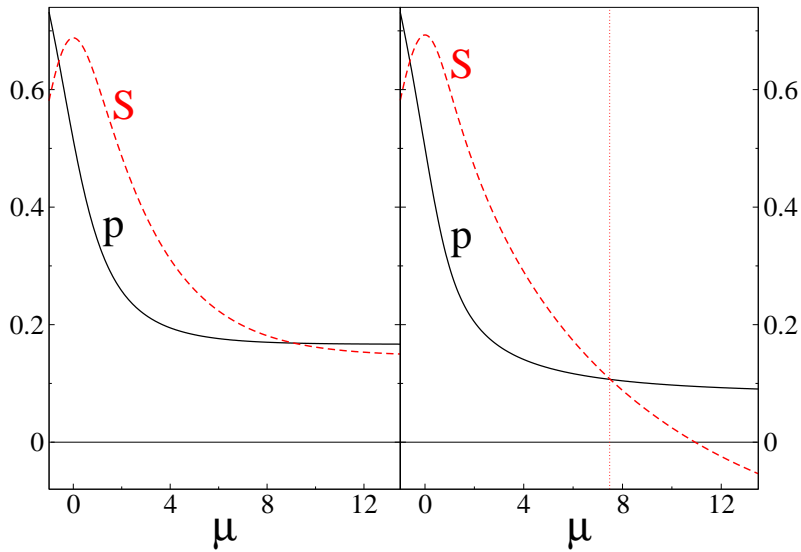}
\end{center}
\caption{Density of active items, $p$, and entropy density, $s = S/N$,
as a function of the chemical potential, $\mu$,
in the RS solution of the hitting set problem for $L=2$ and $K=6$ (left panel)
and for $L=6$ and $K=12$ (right panel).}
\label{fig:rs}
\end{figure}

Belief propagation amounts to using these equations on our problem, even if the graph is not a tree. The equations can be simplified by introducing
 two local cavity fields on each edge of the 
graph, defined as:
$e^{\mu h_{i \to a}} = Z_1^{(i \to a)} / (Z_0^{(i \to a)} + Z_1^{(i \to a)})$, 
and $e^{\mu v_{a \to i}} = Y_1^{(a \to i)} / Y^{(a \to i)}$. They become:
\begin{eqnarray} \label{eq:cavity}
e^{\mu h_{i \to a}} &=& \frac{ \exp (- \mu ) } { \exp( - \mu ) + \exp \left( 
\sum_{b \in \partial i \setminus a} \mu v_{b \to i} \right) }\\
\nonumber
e^{\mu v_{a \to i}} & = & 1 - \prod_{j \in \partial a \setminus i}
\left( 1 - e^{\mu h_{j \to a}} \right).
\end{eqnarray} 
The replica symmetric (RS) solution to this problem amounts to assuming that there is
a unique solution to this equation, and in the case of random regular graph it 
must be a translation invariant solution:
$h_{i \to a} = h_{RS}$ and $v_{a \to i} = v_{RS}$, $\forall (i,a)$, with:
\begin{equation} \label{eq:fRS}
\mu v_{RS} = \ln \left \{ 1 - \left[ \frac{ e^{\mu (L-1) v_{RS} } } 
{ e^{- \mu} 
+ e^{\mu (L-1) v_{RS} } } \right]^{K-1} \right \}.
\end{equation} 
Solving these equations, one can obtain the density of defective items as well
as the entropy of the system using the usual formulas for the RS Bethe free
energy (see for instance~\cite{marc1}). Figure \ref{fig:rs} shows the
results for two values of the degree pairs $L,K$. This shows that the RS
solution fails at high chemical potential and low density of active items, at
least for $L=6$ and $K=12$, because it obtains a negative entropy.

A one step replica symmetry breaking (1RSB) solution can be constructed along
the lines of reference \cite{marc1} (see also~\cite{XOR} for an exactly solvable case). They take a particularly simple form in
the large $\mu$ limit. In this limit, the survey propagation (SP)
equations~\cite{SP,marc2} can be written in terms of one message $u_0^{a \to
  i} $ per edge of the graph. The SP equations for the hitting set problem are~\cite{MezTar}:
\begin{equation} \label{eq:1rsbu0}
u_0^{a \to i} = 1 - \prod_{j \in \partial a \setminus i} \left( \frac{\prod_{b \in \partial j \setminus a}
u_0^{b \to j}}{e^{-y} + \left( 1 - e^{-y} \right ) \prod_{b \in \partial j \setminus a} u_0^{b \to j}} \right).
\end{equation}

\begin{figure}
\begin{center}
\includegraphics[scale=0.35,angle=270]{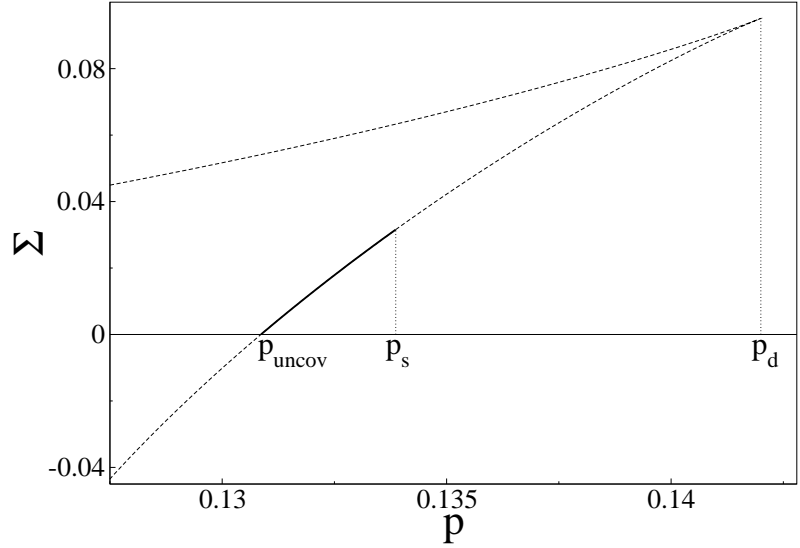}
\end{center}
\caption{Complexity $\Sigma$ as a function of the density
of active variables $p$ for $L=4$ and $K=8$. 
$p_{uncov}$ (where $\Sigma = 0$) is the minimal covering density. Below 
$p_{uncov}$ it is not
possible to find solutions.}
\label{fig:RSB}
\end{figure}

\begin{figure}
\begin{center}
\includegraphics[scale=0.35,angle=270]{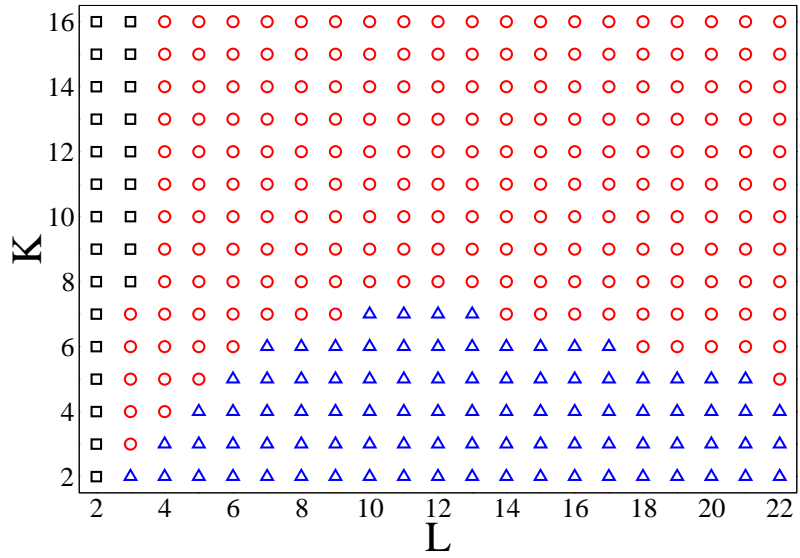}
\end{center}
\caption{Phase diagram of the Hitting set problem.
Squares, circles and triangles correspond, respectively,
to the values of $L$ and $K$ for which the minimal hitting set configurations
are obtained in a RS phase, a 1RSB 
 phase , and a full RSB phase. For the cases RS and 1RSB, we have obtained a closed form expression for the 
value of  the minimal density $p$.
}
\label{fig:PD}
\end{figure}

For random regular graphs, the 1RSB solution can be obtained by assuming that
$u_0^{a \to i}$ is translation invariant, $ u_0^{a \to i}=u_0$. Equation
(\ref{eq:1rsbu0}) can be solved easily, and from this solution one can
compute, using the technique of~\cite{marc1}, the complexity function.
In the present case, this function gives $1/N'$ times the logarithm of the
number of clusters of solutions, versus the optimal density of the cluster
$p$. Figure \ref{fig:RSB} gives the result for the case $L=4$ and $K=8$. The
value of the density where the complexity goes to zero gives the minimal
density such that a hitting set exists. 
The 1RSB solution is stable to further replica symmetry breaking effects for a wide range of values of the degree pairs $L,K$. Figure \ref{fig:PD} summarizes the nature of the
low density phase when one varies $L$ and $K$.
%M
\subsection{Survey propagation and survey inspired decimation}
As proposed first in~\cite{SP}, the equations obtained from the cavity
method can be applied to a single instance of the inference problem, and
turned into efficient algorithms. We have studied in~\cite{MezTar} two
decimation algorithms, one based on the BP equations (\ref{eq:cavity}) and one
based on the SP equations (\ref{eq:1rsbu0}). In both cases the strategy is the
same: one iterates the equations in parallel, starting from a random initial
condition. If a fixed point is reached, one computes the degree of
polarization $p_i$ of each variable. $p_i$ measures to what extent the
marginal probability distribution of variable $i$ is biased, either towards
$x_i=0$ or towards $x_i=1$. In the BP case, $p_i$ is defined as
\begin{equation}
p_i= \left | \frac{ \exp (- \mu ) } { \exp( - \mu ) + \exp \left( 
\sum_{b \in \partial i } \mu v_{b \to i} \right) }-\frac{1}{2}\right |\ .
\end{equation}
In the case of SP, it is defined as
\begin{equation}
p_i = \left| \frac{ \left ( 1 - \prod_{b \in \partial i }
u_{0}^{b \to i} \right ) e^{-y} }
{ e^{-y} + \left(1 - e^{-y} \right) \prod_{b \in \partial i }
u_{0}^{b \to i} }-\frac{1}{2}\right |\ .
\end{equation}

The idea of the BP (or SP) inspired decimation algorithm is to identify the most polarized variable
from BP (or SP), and fix its value $x_i$ to its most probable value. Then variable $x_i$ is 
%C then 
removed from the graph; if $x_i=1$ the tests connected to $i$ are also removed. This procedure is then iterated until all variables are fixed. A subtle issue concerns the values of $\mu$ chosen in BP (resp. the value of $y$ chosen in SP). In order to get a better convergence, we compute the entropy versus $\mu$ in the BP case (resp. the complexity versus $y$ in the SP case), and fix the value of $\mu$ (resp. $y$) to the
largest value such that
the entropy (resp. the complexity) is positive. In this way $\mu$ (resp. $y$) evolves during the decimation procedure. 

In order to get some point of comparison, we have compared the BP and SP inspired decimation to a greedy algorithm, simply defined by the iteration of the procedure: find the  variable $i$ of largest degree, fix it to $x_i=1$, clean the graph.
On one instance of a random regular graph with 
 $L=4,\ K=6,\ N=12288$, these three algorithms have obtained some hitting sets with 
the following  minimal densities:
\begin{itemize}
\item
Greedy algorithm: $p\simeq 0.212$.
\item
 BP inspired decimation: $p\simeq 0.186$. 
\item
SP inspired decimation: $p\simeq 0.182$. 
\end{itemize}
Notice that the prediction from the previous section states that, for $L=4,\ K=6$, 
the minimal density necessary to obtain a hitting set, for an infinite
graph, should be  $p=0.178$. On this example, and in various other experiments that we have tried, 
SP inspired decimation performs slightly better than other algorithms. It would be interesting
to extend such comparisons more systematically.
\section{Perspectives}
Group Testing offers a variety of interesting questions, some of which also have some practical relevance.
One of the results that could turn out to be important is the fact that the message passing 
%C approach
approaches to 
the group testing inference problem seem to be fast and efficient. They are also easily generalizable to 
the case of imperfect tests. One can expect that this will be useful in realistic applications of group testing.

From the point of view of statistical physics, the search for hitting sets gives a new class of problems which 
%C exhibit
exhibits in many cases the general pattern
of 1RSB. In these cases, the hyper-vertex cover problem is thus under much  better control than the usual vertex cover (which exhibits full RSB). Actually, technically these problems are rather simple to solve even at the
1RSB level. They could 
%C thud 
thus offer an interesting practice field to develop
mathematical tools.

%Acknowledgements: Cristina Toninelli, Marco Tarzia ??oppure autori.
We thank  Irina Rish, Greg Sorkin and Lenka Zdeborova for interesting and stimulating discussions. This work has been supported in part by the 'EVERGROW' EC consortium in the FP6-IST program.

%\section*{References}

\end{document}